\documentclass[12pt, a4paper]{article}
\usepackage{amsmath,amssymb,amsfonts}
\usepackage{graphicx}
\usepackage[colorlinks]{hyperref}
\usepackage{makecell}
\usepackage{cite}
\setcellgapes{3pt}

\begin{document}


\title{Dependence of scalar matter vacuum energy, induced by a magnetic topological defect, on the coupling to space-time curvature     \\\phantom{hjbj}}%

\author{V.M. Gorkavenko$^{1,2}$\thanks{Corresponding author. \textit{Email address:} \textbf{gorkavol@knu.ua} (Volodymyr Gorkavenko)}, A.O. Zaporozhchenko$^{1}$, M.S. Tsarenkova$^{1}$ \\
\phantom{jvjjv}\\
\it \small ${}^1$ Faculty of Physics, Taras Shevchenko National University of Kyiv,\\
\it \small 64, Volodymyrs'ka str., Kyiv 01601, Ukraine\\
\it \small ${}^2$ Bogolyubov Institute for Theoretical Physics, National Academy of Sciences of Ukraine,\\
\it \small 14-b, Metrolohichna str., Kyiv 03143, Ukraine}

\date{}

\maketitle

\begin{abstract}
We considered the vacuum polarization of a quantized charged scalar matter field in the background of a topological defect modeled by a finite-thickness tube with magnetic flux inside. The tube is impenetrable for quantum matter, and a generalized boundary condition of the Robin type is imposed at its surface. We have shown that in the flat space-time, the total induced vacuum energy does not depend on the coupling $(\xi)$ of the scalar field's interaction with the space-time curvature, only for the partial cases of the Dirichlet and Neumann boundary conditions on the tube's edge. However, for generalized Robin boundary conditions, the total induced energy depends on the coupling $\xi$ in flat space-time, at least for negative values of the boundary condition parameter $-\pi/2<\theta<0$.

\end{abstract}

\maketitle

\section{Introduction}

After the Casimir paper \cite{Cas}, it became clear that the presence of external boundaries leads to changes in the vacuum energy density. The observed effect of changes in vacuum energy density is the Casimir force. In particular, two conducting plates short distances from each other experience attraction.
Subsequently, this problem was considered for different boundary shapes and materials.
The boundary manifolds are usually chosen as disconnected noncompact objects (such as the infinite plates) and closed compact objects (such as a box or a sphere), see, e.g., \cite{Eli,Most,Bordag1}. 
The importance of such research is indicated by the fact that a thorough study of the Casimir effect can put constraints on theories beyond the Standard Model \cite{MostBSM}.

In this paper, we will be interested in the boundary manifold as a connected, noncompact object. Namely, we will consider an infinite cylindrical tube impenetrable for a matter field with a magnetic flux inside. 
In this case, the Aharonov-Bohm effect \cite{Aha} plays an important role since the magnetic field inside the tube can interact with the quantum matter field outside the tube. 
In the framework of second-quantized theory, vacuum polarization affects the matter field and leads to the induction of a vacuum current and magnetic flux outside the tube. The boundary condition in this setup significantly influences the behavior of the matter field outside the tube. Since the quantum effects, in this case, arise from both the imposition of the boundary condition and the presence of a magnetic field inside the tube, this phenomenon is referred to as the Casimir-Bohm-Aharonov effect \cite{Sit}.

The phenomenon of vacuum polarization near an impenetrable magnetic tube is of significant interest due to its wide range of physical applications.
As a result of phase transitions with spontaneous gauge symmetry breaking in the early Universe or in condensed matter physics, linear topological defects can form. They can be modeled by magnetic tubes of finite radius. The intricate physics inside the topological defect can be modeled by considering the tube impermeable to the matter field and varying the boundary condition on its surface.
From an astrophysical perspective, the model of an impenetrable magnetic tube can be viewed as cosmic strings that could form in the early Universe \cite{Kibble,Vilen81,Vilen95,Hindmarsh}. In this case, the gauge field inside the cosmic string does not have to be an electromagnetic field. 
It is possible that there is a gauge field of the early Universe of some group $U_X(1)$ beyond the Standard Model. In this case, the condition of quantization for the analog of the magnetic flux will be
\begin{equation}\label{quantflux}
    \Phi=\oint d \textbf{x} \textbf{A}=\frac{2\pi n}{e_{cond}},
\end{equation}
where $n\in
\mathbb{Z}$ ($\mathbb{Z}$ is the set of integer numbers), $e_{cond}$ is a coupling between the condensate of the scalar (Higgs) field that generates the cosmic string and the gauge field $U_X(1)$.
In condensed matter physics, the impenetrable magnetic tube can be considered as a model of Abrikosov-Nielsen-Olesen vortex in superconductors of the second group (in this case $e_{cond}=2e$ in \eqref{quantflux} is the charge of Cooper pair and $\Phi$ is a flux of the magnetic field) \cite{Abr,Nielsen} or as disclinations in nanoconical structures of two-dimensional materials like graphene, see, e.g., \cite{Krishnan,HeibergA,Vlasii,Naess,LowTemp,spinor2019}.

It should be noted that initially the Casimir-Bohm-Aharonov effect was considered both for scalar and fermion quantum matter under the assumption that the transverse size of the tube is zero, which corresponds to the singular
magnetic vortex. The advantage of this approach was the ability to carry out all the calculations analytically, whereas its disadvantage was the divergence in the computation of the induced quantum effects due to the zero radius of the tube,
 see, e.g.,  \cite{Sit}, \cite{Ser,Gor,Fle,Par,Si6,Si7,Sit1,BabSit,our1,our2}. As it turned out, the problem of divergences can be eliminated by taking into account the finite size of the space with a magnetic vortex.

In this paper, we will consider the vacuum polarization of a charged massive scalar matter field $\psi$ in the background of a finite transverse size $r_0$ impenetrable for the matter field magnetic tube. In this case,  on the tube's surface, one can impose
generalized boundary conditions of the Robin type
\begin{equation}\label{Robin}
    (\cos \theta\, \psi + \sin \theta\, r \partial_r \psi)|_{r_0} =0,
\end{equation}
where $\theta$ is a parameter of the boundary condition, $-\pi/2 \leq \theta< \pi/2$. Here, cases $\theta=0$ and $\theta=-\pi/2$ correspond to Dirichlet and Neumann boundary conditions, respectively. The induced vacuum energy for the case of the Dirichlet boundary condition on the edge of the finite-size impenetrable magnetic tube was considered in  \cite{our3,our2011,our2013}, for the Neumann boundary condition in \cite{indenerN}. The induced vacuum current and the induced magnetic flux for the case of the Dirichlet boundary condition were considered in \cite{our2016}, and the case of the Neumann boundary condition was considered in \cite{our2022}. The induced magnetic flux for the Robin boundary condition on the tube edge (arbitrary value of the parameter $\theta$) was considered in \cite{our2022prd}.

In this paper, we will consider the vacuum polarization of the charged scalar matter outside the impenetrable finite-thickness magnetic tube with the Robin boundary condition at its edge. While linear topological defects can give rise to a conical spacetime with an angular deficit \cite{Vilen95}, we restrict our attention to a simplified scenario without an angle deficit, treating the space outside the magnetic tube as flat Minkowski spacetime. We will be interested in the dependence of the induced vacuum energy on the coupling to the space-time curvature ($\xi$). An interesting fact is that the dependence of the energy-momentum tensor of the scalar field on the coupling $\xi$ remains even in the case of flat space-time.
    For $(d+1)$-dimensional
space-time only one value of the coupling $\xi$ is highlighted from a theoretical point of view $\xi_c=(d-1)/(4d)$, when the conformal
invariance of the theory is achieved 
\cite{Penrose,Chernikov,CurtisG,Birrell}. 
It was shown in \cite{our2011} that the total induced vacuum energy in flat space-time in the background of the impenetrable magnetic tube with the Dirichlet boundary condition on its edge does not depend on the coupling $\xi$. 
However, the dependence of the total induced vacuum energy of the quantized scalar field in flat space-time on the coupling to the space-time curvature under other boundary conditions has not yet been studied. 

The study of effects caused by the interaction of a scalar field with the curvature of space-time is of considerable interest.
 For instance, the introduction of the direct interaction of the Standard Model Higgs field with gravity influences the stability of our Universe. Value of $\xi$ coupling affects the tunneling time of transition from the electroweak vacuum to the second minimum of the Higgs field, especially when modification of the Standard Model is included, see, e.g., \cite{Bentivegna}.

\section{Energy density}

Lagrangian for a complex scalar massive field $\psi$  in
$(d+1)$-dimensional space-time has form
\begin{equation}\label{0}
\mathcal{L}=({\mbox{$\nabla$}}_\mu\psi)^*({\mbox{$\nabla$}}^\mu\psi)-(m^2+\xi R)\psi^*\psi,
\end{equation}
where ${\mbox{ $\nabla$}}_\mu$ is the covariant derivative involving both affine and
bundle connections and $m$
is the mass of the scalar field, $R^{\mu\nu}$ is the Ricci tensor, $R=g_{\mu\nu}R^{\mu\nu}$ is the
scalar curvature of space-time and $\xi$ is the coupling  of the
scalar field to the scalar curvature of space-time.

The general theory of quantum fields in curved space-time is more complicated compared to the case of Minkowski space-time, see, e.g., \cite{Birrell,Fulling,Grib,Wald,Ford,Parker}.  
Even though the notion of well-defined modes is often problematic in curved spacetime, in the specific case of a static background, the operator of the quantized charged scalar field can be expressed in the form
\begin{equation}\label{a11}
\Psi(x^0,\textbf{x})=\sum\hspace{-1.4em}\int\limits_{\lambda}\frac1{\sqrt{2E_{\lambda}}}
\left[e^{-iE_{\lambda}x^0}\psi_{\lambda}(\textbf{x})\,a_{\lambda}+
  e^{iE_{\lambda}x^0} \psi_\lambda^\ast(\textbf{x})\,b^\dag_{\lambda}\right],
\end{equation}
where $a^\dag_\lambda$ and $a_\lambda$ ($b^\dag_\lambda$ and
$b_\lambda$) are the scalar particle (antiparticle) creation and
annihilation operators; $\lambda$ is the set of parameters (quantum numbers) specifying the state;
  $E_\lambda=E_{-\lambda}>0$ is the energy of the state; symbol
  $\sum\hspace{-1em}\int\limits_\lambda$ denotes summation over discrete and
  integration (with a certain measure) over continuous values of
  $\lambda$; wave functions $\psi_\lambda(\textbf{x})$ are the
  solutions to the stationary equation of motion,
\begin{equation}\label{a12}
 \left\{-{\mbox{\boldmath $\nabla$}}^2  + (m^2+\xi R)\right\}  \psi_\lambda(\textbf{x})=E^2_\lambda\psi(\textbf{x}),
\end{equation}
where $\mbox{\boldmath $\nabla$ }$  is the covariant differential operator in an external (background) field.

In general, the vacuum energy density is determined as the vacuum
expectation value of the time-time component of the energy-momentum
tensor, which is given  by the formal expression \cite{our1,our2,our2011}
\begin{equation}\label{a14}
\varepsilon=\langle \rm
vac|\left[\partial_0\Psi^\dag\partial_0\Psi+\partial_0\Psi\partial_0\Psi^\dag+(1/4-\xi)\mbox{\boldmath
$\nabla$}^2(\Psi^\dag\Psi+\Psi\Psi^\dag)\right]- \xi R^{00}(\Psi^\dag\Psi+\Psi\Psi^\dag)|\rm vac\rangle.
\end{equation}
In the following, we consider the case where
space-time is flat $R=0$, then the vacuum energy density is defined as
\begin{equation}\label{a14a}
\varepsilon=\varepsilon_{can}+(1/4-\xi)\varepsilon_\xi=\sum\hspace{-1.4em}\int\limits_{\lambda}E_\lambda\psi^*_\lambda(\textbf{x})\,\psi_\lambda(\textbf{x})+(1/4-\xi)\mbox{\boldmath
 $\nabla$}^2
   \sum\hspace{-1.4em}\int\limits_{\lambda}E^{-1}_\lambda\psi^*_\lambda(\textbf{x})\,\psi_\lambda(\textbf{x})
\end{equation}
and contains dependence on the $\xi$ parameter. Here, $\varepsilon$ is obtained from the energy-momentum
tensor by varying the action with Lagrangian \eqref{0} with respect to the metric, $\varepsilon_{can}$ corresponds to the canonical energy density $T^{00}_{can}$ from Noether's theorem without taking into account the interaction of the scalar field with the curvature of space-time. As can be seen, at $\xi=1/4$, the canonical relation for the energy density is restored.

We consider a static background in the form of a cylindrically symmetric tube of finite transverse size carrying magnetic flux, and focus on the effects of vacuum polarization outside the tube. Although linear topological defects can lead to the appearance of a conical space with an angle deficit, in this paper, we will consider the simplified case of no angle deficit and consider the space outside the magnetic tube as a Minkowski space. The coordinate system is chosen so that the tube is aligned along the $z$ axis.
  The tube in 3-dimensional space can be naturally generalized to
 the $(d-2)$-tube in $d$-dimensional space by adding extra $d-3$
 dimensions as longitudinal ones.
 The covariant derivative is $\nabla_0=\partial_0$, $\mbox{\boldmath
$\nabla$}=\mbox{\boldmath $\partial$}-{\rm i} \tilde e\, {\bf V}$
with $\tilde e$ being the coupling constant of dimension
$m^{(3-d)/2}$ and the vector potential possessing only one nonvanishing component is given by
\begin{equation}\label{4}
V_\varphi=\Phi/2\pi,
\end{equation}
outside the tube; here,  $\Phi$ is the value of the gauge flux inside
the $(d-2)$-tube and $\varphi$ is the angle in  polar $(r,\varphi)$
coordinates on a plane that is transverse to the tube.

The solution of \eqref{a12} for flat space-time satisfying the boundary conditions of the Robin type \eqref{Robin} outside the impenetrable
tube of  radius $r_0$ takes the form  
\begin{equation}\label{6}
\psi_{kn{\bf p}}({\bf x})=(2\pi)^{(1-d)/2}e^{{\rm i}\bf{p
x}_{d-2}}e^{{\rm i}n\varphi}\Omega_{|n- {\tilde e}
\Phi/2\pi|}(\theta,kr,kr_0),
\end{equation}
where $\theta$ is the parameter of the boundary condition,
\begin{equation}\label{7}
\Omega_\rho(\theta,u,v)=\sin\mu_\rho(\theta,v) J_{\rho}(u)-\cos\mu_\rho(\theta,v) Y_{\rho}(u),
\end{equation}
\begin{align}
    & \sin\mu_\rho(\theta,v)=\frac{\cos\theta\, Y_{\rho}(v)+\sin\theta\, v Y'_{\rho}(v)}{\sqrt{\left[\cos\theta J_{\rho}(v)+\sin\theta\, v J'_{\rho}(v)\right]^2+\left[\cos\theta\, Y_{\rho}(v)+\sin\theta\, v Y'_{\rho}(v)\right]^2}},\\
  &  \cos\mu_\rho(\theta,v)=\frac{\cos\theta\, J_{\rho}(v)+\sin\theta\, v J'_{\rho}(v)}{\sqrt{\left[\cos\theta J_{\rho}(v)+\sin\theta\, v J'_{\rho}(v)\right]^2+\left[\cos\theta\, Y_{\rho}(v)+\sin\theta\, v Y'_{\rho}(v)\right]^2}},
\end{align}
and $0<k<\infty$, $-\infty<p^j<\infty$ ($j=\overline{1,d-2}$), $n\in
\mathbb{Z}$ ($\mathbb{Z}$ is the set of integer numbers),
 $J_\rho(u)$ and $Y_\rho(u)$ are the Bessel functions of order $\rho$ of the first and  second
 kinds, the prime near the function means derivative with respect to the function argument. Solutions \eqref{6} obey orthonormalization condition
\begin{equation}\label{eq8}
\int\limits_{r>r_0} d^{\,d}{\bf x}\, \psi_{kn{\bf p}}^*({\bf
x})\psi_{k'n'{\bf p}'}({\bf
x})= \frac{\delta(k-k')}{k}\,\delta_{n,n'}\,\delta^{d-2}(\bf{p}-\bf{p}').
\end{equation}

Unfortunately, this relation \eqref{a14a} suffers from ultraviolet
divergencies. A well-defined quantity can be obtained by employing regularization (using zeta-function \cite{Dow,Haw} or heat-kernel formalism \cite{Bordag95,Bordag96}) and then renormalization procedures, see, e.g., \cite{Most}. 
However, the good news is that
in our case, the magnetic field configuration is excluded from the matter field region, irrespective of the
number of spatial dimensions, the renormalization procedure is reduced to making one subtraction, namely, to subtract the contribution
corresponding to the absence of the magnetic flux, see \cite{BabSit}.

\section{The case of the $(2+1)$-dimensional space-time}

As it was shown in \cite{our2013},
the case of the vacuum polarization in the presence of an impenetrable magnetic tube for the space-time of arbitrary dimension can be generalized from the case of $2+1$ space-time.  Thus, we shall now restrict ourselves to focusing on the plane $z=0$ which is orthogonal to the vortex, that is, nothing more than the case of $(2+1)$-dimensional space-time. In this case,
 the renormalized vacuum energy density  takes the form
\begin{multline}\label{c2}
\varepsilon_{ren}=\frac1{2\pi}\left\{ \int\limits_0^\infty
  dk\,k\left(k^2+m^2\right)^{1/2}G(\theta,kr,kr_0,\Phi)+\right.\\\left.+(1/4-\xi)\triangle_r\int\limits_0^\infty
  dk\,k\left(k^2+m^2\right)^{-1/2}G(\theta,kr,kr_0,\Phi)\right\},
\end{multline}
where $\triangle_r=\partial^2_r+r^{-1}\partial_r$ is the transverse radial part of the Laplacian, 
\begin{equation}\label{Gfunction}
    G(\theta,kr,kr_0,\Phi)=S(\theta,kr,kr_0,\Phi)-S(\theta,kr,kr_0,0)
\end{equation}
and
\begin{equation}\label{a29a}
S(\theta,kr,kr_0,\Phi)=\sum_{n\in\mathbb
 Z}\Omega^2_{|n- {\tilde e}
\Phi/2\pi|}(\theta,kr,kr_0).
\end{equation}
Because of the infinite range of summation, the $S$-function
will depend only on the fractional part of the flux
\begin{equation}\label{a29a1}
   F=\frac{\tilde e\Phi}{2\pi}-\left[\!\left[\frac{\tilde e\Phi}{2\pi}\right]\!\right],\quad(0\leq F < 1),
\end{equation}
where $[[u]]$ is the integer part of quantity u (i.e. the integer
which is less than or equal to u). So, we get
\begin{equation}\label{a29a}
S(\theta,kr,kr_0,F)= \sum_{n=0}^\infty[\Omega^2_{n+F}(\theta,kr,kr_0)+\Omega^2_{n+1-F}(\theta,kr,kr_0)]
\end{equation}
and conclude that induced vacuum energy density \eqref{c2} depends  on $F$, i.e. it is periodic in the flux $\Phi$ with a
period equal to $2\pi {\tilde e}^{-1}$. Moreover, the induced vacuum energy density value is symmetric under the substitution $F \rightarrow 1-F$.

In the absence of the magnetic flux in the tube $S$-function
takes the form
\begin{equation}\label{c1}
S(\theta,kr,kr_0,0)=\Omega^2_{0}(\theta,kr,kr_0)+ 2\sum_{n=1}^\infty\Omega^2_{n}(\theta,kr,kr_0),
\end{equation}
so one can get
\begin{multline}\label{c11a}
    G(\theta,kr,kr_0,F)=\Omega^2_{F}(\theta,kr,kr_0)+\Omega^2_{1-F}(\theta,kr,kr_0)-\Omega^2_{0}(\theta,kr,kr_0)+\\+\sum_{n=1}^\infty[\Omega^2_{n+F}(\theta,kr,kr_0)+\Omega^2_{n+1-F}(\theta,kr,kr_0)-2\Omega^2_{n}(\theta,kr,kr_0)].
\end{multline}

It should be noted that in practice, to speed up the numerical computation of the $G$-function \eqref{Gfunction}, it is better to extract in the $S$-function terms corresponding to singular magnetic vortex
\begin{equation}\label{a29b}
S(\theta,kr,kr_0,F)=S_{0}(kr,F)+S_{1}(\theta,kr,kr_0,F),
\end{equation}
where $S_{0}(kr)$ corresponds to the appropriate series in the case
of the vacuum polarization by  a singular magnetic
vortex\footnote[1]{In this case only regular solutions of the equation of motion \eqref{a12} are used.} \cite{Sit1,our1,our2} where summation can be performed over infinite limits
\begin{equation}\label{a29b1}
 S_0(kr,F)= \sum_{n=0}^\infty\left[
J^2_{n+F}(kr)+J^2_{n+1-F}(kr)\right] =\int\limits_0^{kr}\!
 d\tau\left[J_F(\tau)J_{-1+F}(\tau)+J_{-F}(\tau)J_{1-F}(\tau)\right].
\end{equation}
In the case of half-integer magnetic flux $F=1/2$, we get
\begin{equation}\label{a29b1a}
 S_0(kr,F=1/2)= \frac{2}{\pi}\int\limits_0^{2kr}\!
 d\tau\frac{\sin\tau}{\tau}=\frac{2}{\pi}{\rm Si}(2kr),
\end{equation}
where ${\rm Si}(z)$ is the sine integral function.
The $ S_1(\theta,kr,kr_0,F)$ in \eqref{a29b} is a correction term due to the finite thickness
of the vortex
\begin{equation}\label{Tila29a}
S_1(\theta,kr,kr_0,F)= \sum_{n=0}^\infty[\bar\Omega^2_{n+F}(\theta,kr,kr_0)+\bar\Omega^2_{n+1-F}(\theta,kr,kr_0)],
\end{equation}
where 
\begin{equation}\label{TildeOmega}
   \bar\Omega^2_\rho(\theta,u,v)=\cos^2\mu_\rho(\theta,v) [Y^2_{\rho}(u)-J^2_{\rho}(u)]-2 \sin\mu_\rho(\theta,v) \cos\mu_\rho(\theta,v) Y_{\rho}(u)J_{\rho}(u).
\end{equation}
In the limit of the infinitely thin tube, we have vanishing correction term $S_1(\theta,kr,kr_0\rightarrow 0,F)=0$.

In the absence of the magnetic flux in the tube, we can again perform summation over infinite limits for the singular vortex
\begin{equation}\label{c1b}
S_0(kr,F=0)=J^2_{0}(kr)+ 2\sum_{n=1}^\infty J^2_{n}(kr)=1,
\end{equation}
and write a correction term due to the finite thickness of the tube
\begin{equation}\label{Tila29aF0}
S_1(\theta,kr,kr_0,F=0)= \bar\Omega^2_{0}(\theta,kr,kr_0)+2\sum_{n=0}^\infty \bar\Omega^2_{n+1}(\theta,kr,kr_0).
\end{equation}

\section{Numerical computations}

Unfortunately, due to the complicated form of one-particle solutions in the case of the finite-thickness magnetic tube \eqref{6}, the computation of the vacuum energy density cannot be done analytically and requires numerical methods. 
 We shall restrict ourselves by considering only half-integer magnetic flux $F=1/2$, where the vacuum polarization effect is maximal for the singular magnetic vortex \cite{Sit,our1}.
  For following numerical computations in the case of $(2+1)$-dimensional space-time, it is better to rewrite \eqref{c2} in the dimensionless form \cite{our2011}
\begin{equation}\label{c3}
r^3\varepsilon_{ren}=\alpha_+(\theta,\lambda,x_0,F)+(1/4-\xi)r^3\triangle_r\left(\frac{\alpha_-(\theta,\lambda,x_0,F)}{r}\right),
\end{equation}
where
\begin{equation}\label{c3ab}
\alpha_\mp(\theta,\lambda,x_0,F)=\frac{1}{2\pi}\int\limits_0^\infty
dz\,z\left[z^2+\left(\frac{x_0}\lambda\right)^2\right]^{\mp1/2}
G(\theta,z,\lambda z,F),
\end{equation}
and $x_0=mr_0$, $\lambda=r_0/r$, $\lambda\in[0,1]$.

It is important to highlight that in the case of the singular magnetic vortex and  $(2+1)$-dimensional space-time, functions $\alpha_\mp$  were calculated analytically in \cite{our2}
\begin{multline}\label{b13}
\alpha_{\mp}^{sing}(mr,F)=\\
-\frac{16\sin(F\pi)}{(4\pi)^{2}\Gamma\left(\pm\frac12\right)} \left( mr\right)^{1 \mp \frac12}  \int\limits_1^\infty \frac{d\upsilon}{\sqrt{\upsilon^2-1}}
\cosh[(2F-1)\, \mathrm{arccosh}\, \upsilon] \upsilon^{\pm\frac12-2}K_{1\mp\frac12}(2mrv)\,.
\end{multline}
In the case of half-integer magnetic flux $F = 1/2$, the $\alpha_+$ function can be written as
\begin{multline}\label{b13singa}
\alpha_{+}^{sing}(mr,F=1/2)=
\frac{m^3}{3\pi^2}\left\{\frac \pi2 - \pi mr
\left[K_0(2mr)L_{-1}(2mr)+K_1(2mr)L_0(2mr)\right]+ \right.\\
\left.+\frac{K_0(2mr)}{2mr}-\left[1-\frac{1}{2
(mr)^{2}}\right]K_1(2mr)\right\},
\end{multline}
where $K_\mu(u)$ and $L_\mu(u)$ are the Macdonald 
and modified Struve functions.

Functions $\alpha_\mp(\theta,\lambda,x_0,F)$ \eqref{c3ab} for the finite-thickness magnetic tube can be computed numerically for different values $mr_0$ and $\lambda$ taking the upper limit of integration as $z_{max}$ and upper limit of summation in \eqref{Tila29a} and \eqref{Tila29aF0} as $N_{max}$, where $z_{max}$ and $N_{max}$ are finite but quite large. The explicit values of $N_{max}$ and $z_{max}$ are defined from the condition the result of computation does not change with sufficient accuracy with the increasing of these parameters, see details in \cite{our3,our2011}. Obtained values of $\alpha_\pm$ functions at different $\lambda$ are interpolated and we get the function of parameters $\alpha_\mp(\theta,x,x_0,F)$. 

\begin{figure}[t]
    \centering
    \includegraphics[width=0.55\textwidth]{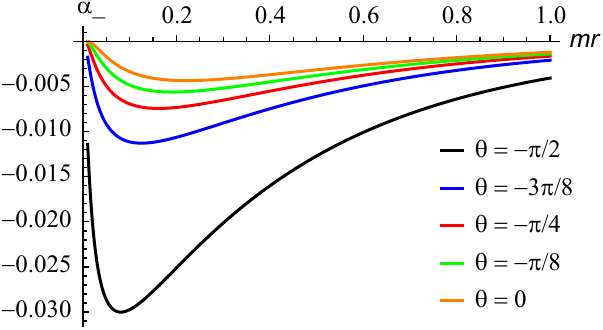}
    \caption{$\alpha_-$ function for fixed thickness of the impenetrable magnetic tube $mr_0=10^{-2}$, half-integer magnetic flux $F=1/2$ and parameter of the boundary condition $\theta=-\pi/2$, $-3\pi/8$, $-\pi/4$, $-\pi/8$, $0$ from down to up line correspondingly. }
    \label{Fig1}
\end{figure}

\begin{figure}[b]
    \centering
    \includegraphics[width=0.95\textwidth]{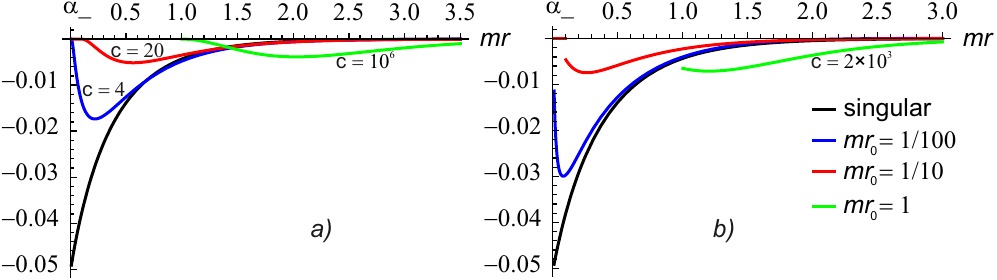}
    \caption{$\alpha_-$ function for the case  of \textit{a}) Dirichlet $(\theta=0)$ and \textit{b}) Neumann $(\theta=-\pi/2)$ boundary condition for the impenetrable magnetic tube of different thicknesses for $mr>mr_0$ and half-integer magnetic flux $F=1/2$.}
    \label{Fig2}
\end{figure}

Since we are interested in $\xi$-dependence of the induced vacuum energy outside the impenetrable magnetic tube, we compute the $\alpha_-$ function for different negative values of the parameter $\theta$ of the Robin boundary condition at its edge as a function of the dimensionless distance from the center of the tube $(x=mr)$. See results in Fig.\ref{Fig1} for the tube of fixed thickness $(x_0=mr_0)$ for different values of parameter $\theta $ and in Fig.\ref{Fig2} for different thickness of the tube for the case of Dirichlet and Neumann boundary conditions on its edge.

\begin{figure}[t]
    \centering
    \includegraphics[width=0.65\textwidth]{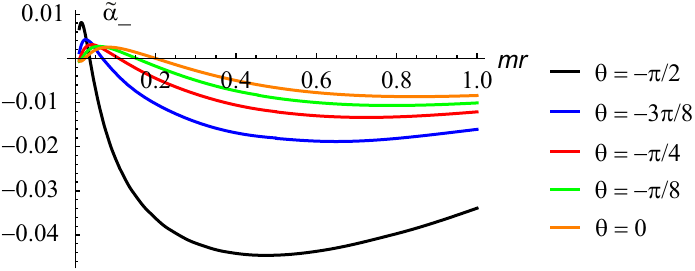}
    \caption{$\tilde \alpha_-$ function for  the impenetrable magnetic tube of thickness $mr_0=10^{-2}$, half-integer magnetic flux $F=1/2$, and different values of parameter of the boundary condition $\theta=-\pi/2$, $-3\pi/8$, $-\pi/4$, $-\pi/8$, $0$ on its edge. }
    \label{Fig3}
\end{figure}

Following \cite{our2011} it is useful to consider the dimensionless function
\begin{multline}\label{m2}
\tilde
\alpha_-(\theta,x,x_0,F)=r^3\triangle_r\left(\frac{\alpha_-(\theta,x,x_0,F)}{r}\right)=\\ \alpha_-(\theta,x,x_0,F)-x\frac{\partial
\alpha_-(\theta,x,x_0,F)}{\partial x}+x^2\frac{\partial^2
\alpha_-(\theta,x,x_0,F)}{\partial x^2},
\end{multline}
and write induced vacuum energy density by the impenetrable magnetic tube as
\begin{equation}\label{m3}
r^3\varepsilon_{ren}=\alpha_+(\theta,x,x_0,F)+(1/4-\xi)\tilde\alpha_-(\theta,x,x_0,F).
\end{equation}
The result of the numerical computation $\tilde \alpha_-$ function is presented in Fig.\ref{Fig3}.

It should be noted that in the case of the singular magnetic vortex and  $(2+1)$-dimensional space-time, the $\tilde \alpha_-$ function  was analytically calculated in \cite{our2}
\begin{multline}\label{b13tilde}
\tilde\alpha_{-}^{sing}(mr,F)=
-\frac{32\sin(F\pi)}{(4\pi)^{2}\Gamma\left(\frac12\right)} \left( mr\right)^{ \frac32}\times \\  \int\limits_1^\infty \frac{d\upsilon}{\sqrt{\upsilon^2-1}}
\cosh[(2F-1)\, \mathrm{arccosh}\, \upsilon] \upsilon^{-\frac12}\left[2K_{\frac32}(2mrv)-2mrvK_{\frac52}(2mrv)\right]\,.
\end{multline}
In the case of half-integer magnetic flux $F=1/2$ this function can be written as 
\begin{equation}\label{b13tilde12}
\tilde\alpha_{-}^{sing}(mr,F=1/2)=-\frac{mr}{\pi^2}( 2 mr K_0(2mr) + K_1(2 mr)).
\end{equation}

\begin{figure}[t!]
    \centering
    \includegraphics[width=0.95\textwidth]{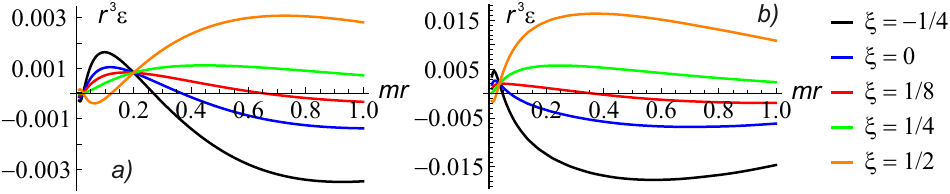}
    \caption{The induced vacuum energy density $r^3 \varepsilon_{ren}$ for different values of coupling scalar field to the space-time curvature $\xi$ for the case of \textit{a})  Dirichlet  $(\theta=0)$ and \textit{b}) Neumann  $(\theta=-\pi/2)$ boundary conditions on the edge of the magnetic impenetrable tube of thickness $mr_0=10^{-2}$  and half-integer magnetic flux $F=1/2$.}
    \label{Fig4}
\end{figure}

The $\alpha_+$ function \eqref{c3ab} represents the induced canonical energy density and was obtained for the case of the impenetrable magnetic tube in \cite{our2011,indenerN} for the case of Dirichlet and Neumann boundary conditions on its edge.
Using these results, we present, as an example, the induced dimensionless vacuum energy density $r^3 \varepsilon_{ren}(mr,\xi)$ for different values of coupling scalar field to the space-time curvature $\xi$ for these cases of boundary conditions on the edge of the magnetic impenetrable tube, see Fig.\ref{Fig4}. In the plot, we construct induced vacuum energy density in the vicinity of the $\xi_c=1/8$, when the conformal
invariance of the theory is achieved. 

One can see that coupling $\xi$ significantly impacts the induced energy density. It is interesting to note that the behaviors of the induced vacuum energy density for the case of Dirichlet and Neumann boundary conditions are similar at large distances from the tube, but different in the vicinity of the impenetrable magnetic tube. In the case of the Dirichlet boundary condition, the corresponding function has two nodal points, whereas in the case of the Neumann boundary condition, the corresponding function has only one nodal point.

 \section{Total induced vacuum energy }

As is well known \cite{Sit}, in the case of the singular magnetic vortex, the total induced vacuum energy is a nonphysical characteristic because of its divergence during the integration of the energy density over radius at small values of the variable. As it was shown in \cite{Sit} and can be seen directly using \eqref{b13singa}, \eqref{b13tilde12}, the induced vacuum energy is proportional to $r_0^{-1}$, where $r_0$ is the lower limit of integration ($r_0\rightarrow0$). It is one of the motivations to consider a magnetic vortex of finite thickness.

The total induced vacuum energy by the impenetrable finite thickness magnetic tube in $(2+1)$-dimensional space-time is well-defined and given by \vspace{-1em}
\begin{multline}\label{m4}
E=\int\limits_{0}^{2\pi}d\varphi\int\limits_{r_0}^\infty
\varepsilon_{ren} \,r dr=E_{can}+(1/4-\xi)E_\xi=\\=2\pi m\left[ \int\limits_{x_0}^\infty
\frac{\alpha_+(\theta,x,x_0,F)}{x^2}\, dx +(1/4-\xi)\int\limits_{x_0}^\infty
\frac{\tilde\alpha_-(\theta,x,x_0,F)}{x^2}\, dx\right],
\end{multline}
where $E_{can}$ corresponds to the canonical definition of the time-time component of the energy-momentum tensor.

In this section, we will not be interested in the computation of $E_{can}$. 
Instead, we will be interested in whether the total induced vacuum energy in flat space-time in the background of the impenetrable magnetic tube depends on the coupling $(\xi)$ of the scalar field's interaction with the space-time curvature. 
To answer this question, we need to know the value of the integral over the $\tilde\alpha_-$ function \eqref{m2}.
This integral  can be
taken by parts
\begin{equation}\label{m5}
E_\xi=2\pi m\int\limits_{x_0}^\infty \frac{\tilde\alpha_-(\theta,x,x_0,F)}{x^2}\,
dx=-2 \pi m x\left.\frac{\partial}{\partial x}\frac{\alpha_-(\theta,x,x_0,F)}{x}\right|_{x=x_0}.
\end{equation}

\begin{figure}[h!]
    \centering
    \includegraphics[width=0.45\textwidth]{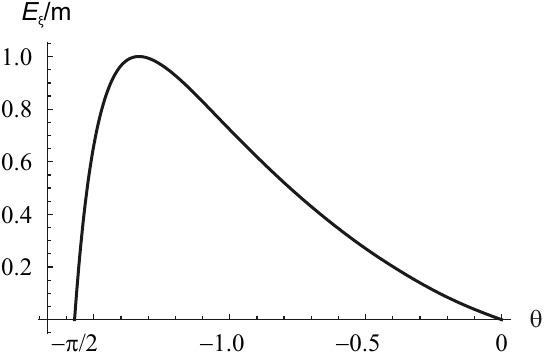}
    \caption{The total induced dimensionless vacuum energy $E_\xi/m$ \eqref{m4} as the function of parameter $\theta$ of Robin boundary conditions on the edge of the magnetic impenetrable tube of thickness $mr_0=10^{-2}$ in the interval $-\pi/2 \leq\theta\leq 0$.}
    \label{Fig5}
\end{figure}

In the case of the Dirichlet boundary condition on the tube edge, $E_\xi$ is zero, see \cite{our2011}. 
Using numerical calculations, we have obtained that induced vacuum energy $ E_xi$ is also zero for the Neumann boundary condition.
However, we have obtained that the induced vacuum energy $E_\xi$ is non-zero in the general case of the Robin boundary conditions for the case of $-\pi/2<\theta<0$. The result of the numerical computation of $E_\xi$ is presented in Fig.\ref{Fig5} for a partial case of the tube thickness $mr_0=10^{-2}$. As one can see, induced vacuum energy $E_\xi$ is zero at the boundaries of the interval $-\pi/2<\theta<0$  and is a positive function within this interval.

The case of positive values of the parameter of boundary condition $\theta$ requires additional careful consideration. In this case, the bound state solutions to the Fock-Klein-Gordon equation will contribute to the vacuum polarization that can cause non-trivial behavior of vacuum effects as in the case of induced magnetic flux \cite{our2022prd}.

\section{Summary} \vspace{-0.5em}

In flat $(d+1)$-dimensional space-time, we obtained general relations for the vacuum polarization of a quantized charged scalar field in the background of a static magnetic field confined within a $(d-1)$-dimensional tube. We considered the general relation for the induced energy density and took into account its dependence on coupling $(\xi)$ of the scalar field interaction with the space-time curvature. 
We assumed that the scalar field cannot penetrate inside the tube and obeys generalized boundary conditions of the Robin type \eqref{Robin} at its surface.  This boundary condition is characterized by one parameter $-\pi/2 \leq \theta< \pi/2$. In partial cases, this boundary condition corresponds to Dirichlet ($\theta=0$) and Neumann ($\theta=-\pi/2$) boundary conditions.
The physically interesting case of $d=3$ corresponds to an infinitely long tube, while the case of $d=2$ corresponds to a ring  and can be applied as a model for describing topological defects in the early Universe or in condensed matter physics.

We showed that the induced vacuum energy, in this general boundary condition case,
\eqref{a29a} depends periodically on the magnetic  
flux inside the tube with a period equal to $2\pi {\tilde e}^{-1}$.  Thus the induced vacuum energy depends only on the fractional part of the magnetic flux. When the magnetic flux satisfies $\Phi=2\pi n {\tilde e}^{-1} $, where  $n\in \mathbb{Z}$, the vacuum polarization effect vanishes. Such behavior reveals the Casimir-Bohm-Aharonov effect \cite{Sit} and is caused by the condition of the impenetrability of the matter field to the region with the magnetic field. It should be noted that in the case when quantized matter penetrates the region with a magnetic field, the vacuum polarization effect is determined by the total magnetic flux in the tube \cite{Cangemi,Fry,Dunne,Bordag,Scandurra,Langfeld,Graham}.

In the simplest case of $(2+1)$-dimensional space-time and half-integer magnetic flux $F=1/2$, we compute term $\tilde\alpha_-$ of the induced energy density proportional to $(1/4-\xi)$ \eqref{m3}.  
The comparison of the vacuum polarization term $\tilde\alpha_-$ with different values of the parameter of boundary condition $\theta$ is presented in Fig.\ref{Fig3}. For the negative values of the parameter $\theta$, one can see that, for tubes of the same thickness, the vacuum polarization effect is the largest in amplitude in the case of the Neumann boundary condition $(\theta=-\pi/2)$ and decreases as $\theta$ increases. The smallest effect occurs in the case of the Dirichlet boundary condition ($\theta=0$). This result is in agreement with the result of \cite{our2022prd} for the induced magnetic flux. Additionally, we have shown that the vacuum polarization effect decreases as the tube thickness increases, see Fig.\ref{Fig2}. 

The case of positive values of the boundary condition parameter $\theta$ requires further careful analysis. In this scenario, the bound state solutions to the Fock-Klein-Gordon equation will contribute to the vacuum polarization, potentially leading to non-trivial behavior of vacuum effects as in the case of induced magnetic flux \cite{our2022prd}.

For the case of Dirichlet and Neumann boundary conditions on the impenetrable magnetic tube's edge, we construct the induced vacuum energy density of the scalar field for different values of coupling $(\xi)$ of the scalar field's interaction with the space-time curvature in the vicinity of the $\xi_c=1/8$,
when the conformal
invariance of the theory is achieved. It is interesting to note that the behaviors of the induced vacuum energy density for the case of Dirichlet and Neumann boundary conditions are similar at large distances from the tube, but different in the vicinity of the impenetrable magnetic tube. In the case of the Dirichlet boundary condition, the corresponding function has two nodal points, but in the case of the Neumann boundary condition, the corresponding function has only one nodal point, see Fig.\ref{Fig4}.

For the case of ($2+1$)-dimensional space-time, we have directly shown that total induced vacuum energy in flat space-time does not depend on the coupling $(\xi)$ of the interaction of the scalar field with the space-time curvature only for the partial cases of Dirichlet and Neumann boundary conditions on the tube's edge. However, for generalized Robin boundary conditions, the total induced energy depends on the coupling $\xi$ in flat space-time, at least for negative values of the boundary condition parameter $-\pi/2<\theta<0$.  The dependence of induced energy term $E_\xi$ \eqref{m4}, proportional to $(1/4-\xi)$,  on the parameter of boundary condition $\theta$ is presented in Fig.\ref{Fig5}.

Since the total induced vacuum energy in the presence of an impenetrable magnetic tube for a space-time of arbitrary dimension can be generalized
from the case of $(2 + 1)$-dimensional space-time \cite{our2013}, we conclude that the total induced vacuum energy in flat space-time
will also depend on the parameter $\xi$ for a space-time of higher dimensions. This question deserves further in-depth study.

\vspace{-1em}

\section*{Acknowledgments}

The work of V.M.G. and M.S.Ts. was supported by Department of target training of
Taras Shevchenko National University of Kyiv and the NAS of Ukraine,
grant No.6$\Phi$-2024. \vspace{-1em}


\end{document}